# Transient infrared nanoscopy resolves the millisecond photoswitching dynamics of single lipid vesicles in water

T. Gölz[1], E. Baù[1], J. Zhang[2], K. Kaltenecker[1,3], D. Trauner[4], S. A. Maier[5,6], F. Keilmann[1*], T. Lohmueller[2*], A. Tittl[1*]

1. Chair in Hybrid Nanosystems, Nano-Institute Munich, Department of Physics, Ludwig-Maximilians-Universität München, 80539 Munich, Germany

2. Chair for Photonics and Optoelectronics, Nano-Institute Munich, Department of Physics, Ludwig-Maximilians-Universität München, 80539 Munich, Germany

3. Attocube Systems AG, 85540 Haar, Germany

4. Department of Chemistry, University of Pennsylvania, Philadelphia, Pennsylvania 19104-6323, United States

5. School of Physics and Astronomy, Monash University, Clayton, Victoria 3800, AUS

6. Department of Physics, Imperial College London, London SW7 2AZ, UK

Email: fritz.keilmann@lmu.de, t.lohmueller@lmu.de, andreas.tittl@physik.uni-muenchen.de

## Abstract

**Understanding the biophysical and biochemical properties of molecular nanocarriers under physiological conditions and with minimal interference is crucial for advancing nanomedicine, photopharmacology[1], drug delivery[2], nanotheranostics[3] and synthetic biology[4]. Yet, analytical methods struggle to combine precise chemical imaging and measurements without perturbative labeling. This challenge is exemplified for azobenzene-based photoswitchable lipids, which are intriguing reagents for controlling nanocarrier properties on fast timescales[5], enabling, for example, precise light-induced drug release processes[2]. Here, we leverage the chemical recognition and high spatio-temporal resolution of scattering-type scanning near-field optical microscopy (s-SNOM)[6,7] to demonstrate non-destructive, label-free mid-infrared (MIR) imaging and spectroscopy of photoswitchable liposomes below the diffraction limit and the tracking of their dynamics down to 50 ms resolution. The vesicles are adsorbed on an ultrathin 10-nm SiN membrane[8], which separates the sample space from the tip space for stable and hour-long observations. By implementing a transient nanoscopy approach, we accurately**

**resolve, for the first time, photoinduced changes in both the shape and the MIR spectral signature of individual vesicles and reveal abrupt change dynamics of the underlying photoisomerization process. Our findings highlight the method's potential for future studies on the complex dynamics of unlabeled nanoscale soft matter, as well as, in a broader context, for host-guest systems[9], energy materials[10,11] or drugs[12,13].**

**Keywords**

infrared spectroscopy, near-field microscopy, s-SNOM, lipid vesicles, photolipids, photoswitches, azo-PC

**Introduction**

Lipid-based nanocarriers, such as liposomes and lipid nanoparticles (LNPs), have emerged as a leading platform technology in nanomedicine[14,15]. Their key advantage lies in the ability to encapsulate hydrophobic drugs or molecular nanoagents for targeted delivery. Unilamellar vesicles, the most basic form of nanocarriers, have already found their way into clinical applications[16]. In addition, LNPs represent arguably the most advanced nanocarrier technology and have played a crucial role for the successful *in vivo* administration of mRNA-based vaccines[17].

Enhancing the performance of liposomal nanocarriers, including both liposomes and LNPs, depends on optimizing strategies to control site-specific release mechanisms and upon an external trigger[18]. Light is ideally suited for this task due to its contactless nature and ease of focus. Photoswitchable molecules integrated into or forming part of the lipid membrane can facilitate liposome release. Recently, Chander *et al.* made significant progress towards this goal by using the azobenzene-based photoswitchable-phosphatidylcholines azo-PC[19], and red-Azo-PC[20], in lipid nanoparticle formulations, enabling controlled drug release upon irradiation at specific wavelengths[2]. Their work demonstrates the feasibility of integrating photolipids into clinically approved lipid formulations, showing large promise for future development.

Imaging and spectroscopy techniques in the mid-infrared (MIR) spectral range are an ideal toolkit for investigating the chemical composition of different organic and inorganic samples[21] due to the wavelength-specific absorption of infrared light by the chemical material's bonds, often referred to as the "spectroscopic fingerprint". In the case of photoswitchable lipid membranes, MIR spectroscopy is particularly useful for analyzing the isomerization of the membrane-embedded photolipids in a label-free and non-destructive manner without interfering with the switching process itself[22,23]. However, conducting MIR imaging and spectroscopy on a single lipid vesicle requires a methodology that simultaneously provides sufficient nanoscale spatial resolution and high temporal resolution to resolve the photoswitching dynamics. Scattering-type scanning near-field microscopy (s-SNOM)[6,7] is ideally suited for this task. In s-SNOM, a laser beam is focused onto the apex of a sharp metallic AFM tip creating highly confined evanescent fields around the apex that yield spatial resolutions down to 20 nm. The method has been highly successfully for studying dried single biological macromolecules[24] and dried lipid monolayers[25], and has already been extended to observe living biological entities in their native environment[8].

A critical gap in optimizing photolipid-nanocarriers has been the lack of effective tools for studying membrane photoisomerization at the single liposome level. Previous studies have shown that azobenzenes quench fluorescence and dye molecules further interfere with the isomerization process[26], which renders fluorescence-based methods less favorable. While atomic force microscopy (AFM)[27], interferometric scattering microscopy[28] and transmission electron microscopy (TEM)[29] allow for investigating liposome shape and size with sufficient resolution, they do not provide any chemical insights into the isomerization process. Nanophotonic based sensing approaches show great promise for spectroscopically tracking complex dynamics, but lack the simultaneously flexibility to image the system[30–33].

Here, we demonstrate the use of in-situ nanoscopy to image and spectroscopically analyze individual photoswitchable lipid vesicles with sizes down to 176 nm in aqueous environments. In contrast to previous investigations on naturally progressing biological systems[8], we present the first in-situ s-SNOM study on actively induced dynamic processes by reversibly changing the morphology of a vesicle through repeated ultraviolet/blue light illumination and tracking its spectral response at 50 ms temporal resolution. We demonstrate not only the possibility to detect and distinguish two photoisomeric states of the lipid molecules on the single lipid vesicle level based on subtle changes in their near-field MIR spectra, but also monitor the photoinduced transformations of the lipids in their aqueous environment in real time.

**In-situ near-field MIR imaging and spectroscopy of a photoswitchable lipid vesicle**

In our experiments, the MIR near field of an irradiated metallic s-SNOM tip probes a water-suspended lipid vesicle through a 10 nm thick SiN membrane, where the vesicle remains adsorbed by van der Waals forces for extended periods of time (**Fig. 1a**). The advantage of using the membrane is that it prevents sample evaporation while also protecting the tip from contamination by the solution (for experimental details, see **Materials and Methods**).

Furthermore, the SiN membrane-based in-situ technique enables hours long mechanically stable s-SNOM measurements, without the need to realign optics even when changing samples[8].

The MIR laser beam is tightly focused by a paraboloidal mirror onto the tip to generate a highly concentrated near field under the apex (**Fig. 1a**, red area). The optical near field penetrates through the SiN layer and probes more than 100 nm into the underlying liquid compartment[8]. Back-scattered coherent MIR light is detected in a Michelson interferometer setup, which allows for the extraction of both the MIR amplitude $s_2$ and phase $\varphi_2$ (**Fig. 1b** and **Materials and Methods**). The choice of (i) a tunable monochromatic MIR laser or (ii) a wideband coherent MIR supercontinuum source allows for the assessment of either (i) the sample's response at a selected molecular vibration with high speed or (ii) the sample's full molecular fingerprint in a matter of seconds to minutes. The latter modality is known as nano-FTIR[7,34]. An additional mode of MIR operation is the "white light" recording of the detector signal, where all spectral components of the wideband continuum contribute, forming a spectrally averaged infrared signal. The reversible photoisomerization is induced by light from either of two LEDs emitting at 365 nm and 465 nm, respectively, which are aligned to illuminate the whole SiN membrane area homogeneously without generating enhanced near fields (**Fig. 1a**, blue area, **Fig. 1b**).

The vesicles under investigation are composed of a 50:50 % mixture of 1,2-Dioleoyl-sn-glycero-3-phosphocholine (DOPC) and azo-PC (see **Fig. 1c** and **Materials and Methods** for details on sample preparation)[19]. The switching wavelengths were chosen from the azo-PC UV-VIS spectra (**Fig. 1d**). Illumination of the lipid vesicles with a wavelength of 365 nm triggers the isomerization of the azobenzene moiety from *trans* to *cis*, while, conversely, illumination at 465 nm switches the molecules to the thermodynamically more stable *trans*-state.

The photoswitching of azo-PC lipids has been analysed by bond-sensitive MIR spectroscopy[22,23], as we confirm for our samples using ATR-FTIR spectra of dried azo-PC films (**Fig. S1**). The carbonyl band at 1743 cm$^{-1}$ remains unaffected by the photoswitching[19,22], whereas clear differences in spectral intensity between the *cis/trans* isomers are evident from the resonances at 1606 cm$^{-1}$, 1495 cm$^{-1}$ and 1463 cm$^{-1}$, which can be assigned to the ring breathing mode of the *trans*-azo-group and CH$_2$-backbone modes found in the azo-PC lipids[23]. All studied resonances feature increased absorption in the *trans*-conformation[23]. In contrast, DOPC lipid ATR-FTIR spectra do not change in intensity with different illuminations and do not show a band at around 1600 cm$^{-1}$ (**Fig. S1**).

The in-situ s-SNOM technique enables MIR spectroscopy on individual lipid vesicles in an aqueous environment at length scales impossible to reach by standard far-field spectroscopy approaches. Since the most interesting lipid resonance at 1606 cm$^{-1}$ would be masked by a strong H$_2$O vibration at 1650 cm$^{-1}$, we suspended all vesicles in this work in D$_2$O, a common practice in FTIR spectroscopy of organic materials[8,35] (see **Fig. S2** for spectra of vesicles in H$_2$O).

We first present a spectrally averaged MIR near-field amplitude image (**Fig. 1e**) ("second demodulation" $s_2$, see **Materials and Methods**) to identify and target a membrane-fixed vesicle in aqueous solution for further spectroscopic measurements.

We subsequently recorded nano-FTIR phase spectra $\varphi_2$ at the vesicle's center (**Figs. 1e** red cross and **1f**), where the highest scattered white-light signal is observed and therefore a high signal-to-noise ratio (SNR) of the recorded spectra is expected. In both photoisomerization states, these spectra show the known characteristic resonances of the lipid system, such as the carbonyl peak, the CH$_2$-backbone modes[22,23], and a resonance at 1606 cm$^{-1}$ (see **Fig. S2** for a detailed assignment to the molecular bonds of DOPC and azo-PC). Notably, the resonance at 1606 cm$^{-1}$ decreases in intensity after illumination at 365 nm, indicative for *trans*-to-*cis* isomerization. In addition, the CH$_2$-backbone modes at 1463 and 1495 cm$^{-1}$ also decrease in intensity. These differences are consistent with our measurements of ensemble averaged far-field ATR-FTIR spectra (**Fig. S1**) and with reports in literature[22,23]. Therefore, the spectra demonstrate our capability to analyze the chemical composition and distinguish between photoisomers of a nanoscale lipid vesicle by their associated nano-FTIR spectra. Note that the spectroscopic signal of a lipid vesicle in H$_2$O outside the 1650 cm$^{-1}$ H$_2$O band (**Fig. S2**) is of similar good quality and therefore should allow future studies in H$_2$O suspension.

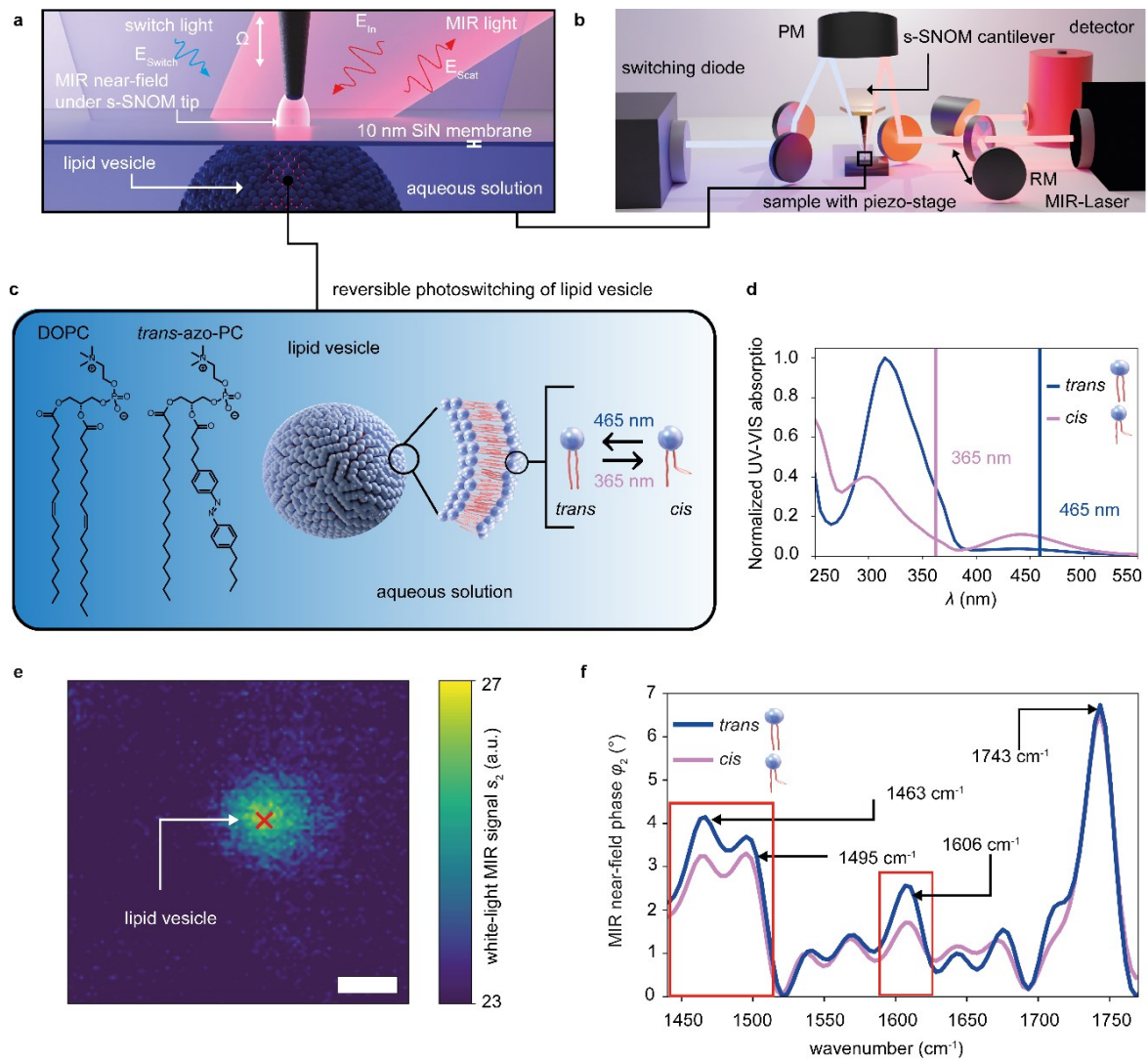

**Figure 1: In-situ s-SNOM infrared spectroscopy of photoactive lipid vesicles. (a)** Operating principle of the membrane-based in-situ s-SNOM method. An s-SNOM tip with its associated near field scans in tapping mode at a tapping frequency Ω above a 10 nm thin SiN membrane, which separates the tip from lipid vesicles suspended in an aqueous medium. The tip and membrane are illuminated by two different light sources: a MIR beam ($E_{in}$) for near-field spectroscopy and imaging and a UV-VIS source ($E_{switch}$) to switch the lipid vesicles between their different photoisomeric states. **(b)** S-SNOM setup with the MIR beam and the UV light focused onto the s-SNOM tip and sample by a parabolic mirror (PM). The focused MIR beam creates an enhanced near field that interacts with the sample underneath. The MIR light backscattered from the tip ($E_{scat}$) containing the spectroscopic information of the liquid sample is collimated and interferes with a reference beam that is reflected by a movable reference mirror (RM). The resulting signal is recorded by a fast response infrared detector **(s. Materials and Methods)**. **(c)** Molecular structure of the DOPC and *trans*-azo-PC lipids constituting the lipid vesicle in a ratio 50:50 and sketch of the light-induced conformational change. **(d)** UV-VIS spectrum of both lipid isomers with switching wavelengths labelled by the blue and violet vertical lines. **(e)** Spectrally averaged MIR amplitude image ($s_2$) of a lipid vesicle in *trans*-state in $D_2O$, scale bar 500 nm. **(f)** MIR near-field phase spectra ($\varphi_2$) of a *trans*- (blue) and *cis*-state (violet) lipid vesicle **(e)** recorded at the position of the red cross. Red boxes highlight two lipid vibrational MIR bands that respond strongly to the switching light.

Based on the nano-FTIR spectra recorded on the lipid vesicle, we chose the intense carbonyl resonance to record resonance-specific MIR images. A larger area scan (15 μm x 15 μm) demonstrates the side-by-side coexistence of numerous vesicles of varying sizes, simultaneously measured in both amplitude $s_2$ (**Fig. 2a**) and phase $\varphi_2$ (**Fig. 2b**). Individual vesicles can be clearly identified and localized, as exemplified by close-up views of four differently sized vesicles, indicated as colored boxes (**Fig. 2c**). The vesicles typically exhibit near-uniform phase $\varphi_2 > 25°$ throughout their inside (shown in red), surrounded by a fringe of around 100 nm width (shown in white).

We extract quantitative phase profiles along the lines indicated in **Fig. 2c** and determine each vesicle's apparent full width at half maximum (FWHM, **Fig. 2d**). Despite only being 176 nm in width, the smallest analysed lipid vesicle is still well resolved (albeit with reduced phase signal), which is consistent with the spatial resolution of our setup on the order of 100 to 150 nm[8]. Importantly, this result demonstrates the capability of our method to characterize nanoscale objects even in water, at length scales way beyond the reach of standard fluorescence and phase contrast microscopies.

The uniform phase within and amongst differently sized vesicles (**Fig. 2b**) indicates that they undergo a deformation and flattening when adhering to the SiN membrane, as sketched in **Fig. 1a**. This interpretation of the measured, uniform signal is in accordance with previous subsurface s-SNOM studies, which showed that the near-field sensing reaches to around 100 nm depth, whereas objects beyond 200 nm below the tip remain virtually invisible[36,37].

To corroborate this flattening behavior, we compare phase profiles of the 660 nm FWHM vesicle and the largest vesicle (next to the brown box in **Fig. 2b**) with an analytical model for the phase signal of a nondeformable polymeric sphere hanging from a single adhesion point on the SiN membrane. We find that the theoretical profiles have a different form and are narrower than the experimental profiles (**Figs. S3a, b**), supporting our hypothesis of vesicle flattening. This observation demonstrates our *optical* methods ability to study details of nanoscale adhesion of soft matter systems in aqueous environments.

Independent further information about the vesicles can be gained by examining the simultaneously acquired *mechanical* images or vesicle "footprints"[8], which show a displacement of the SiN membrane upwards by about 1 nm for the green-boxed vesicle location and up to 2 nm for the largest vesicle (**Figs. S3c, d**). This observation clearly shows that even nanoscale soft-matter objects can be detected via distinct deformations of the SiN membrane, in agreement with previous studies[8,37], and should be applicable to characterize even more complex lipid systems.

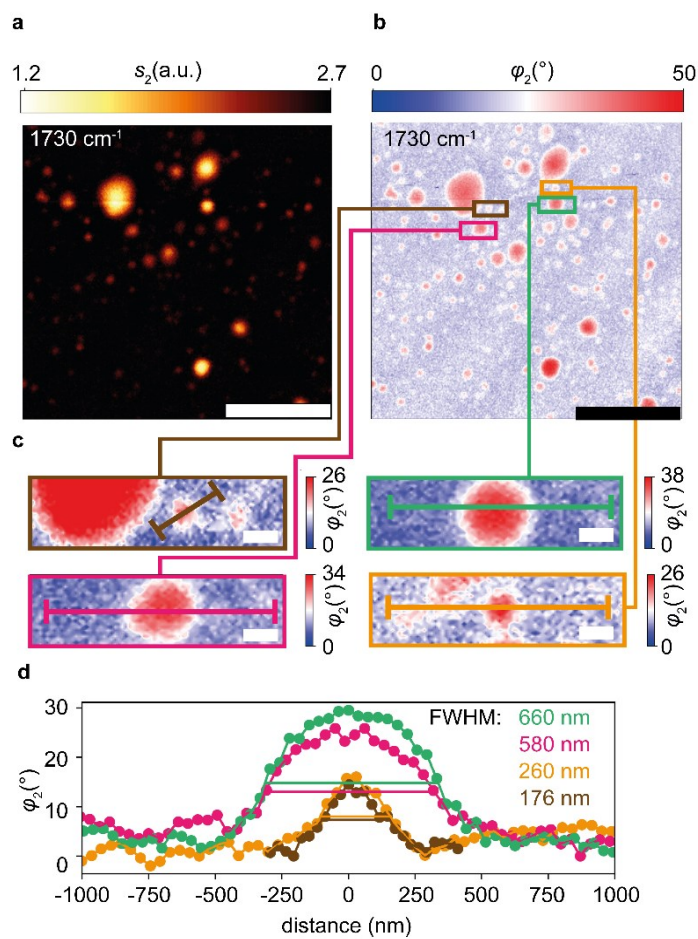

**Figure 2: Chemically specific MIR near-field imaging of nanoscale lipid vesicles**. S-SNOM optical amplitude ($s_2$) **(a)** and phase images ($\varphi_2$) **(b)** of several lipid vesicles suspended in $D_2O$ measured at 1730 cm$^{-1}$ at the carbonyl resonance (see **Fig. 1f**), scale bars 5 µm, acquisition duration 12 min. **(c)** Enlarged phase images ($\varphi_2$) of four differently sized lipid vesicles, scale bars 300 nm. **(d)** Extracted profiles along lines drawn in **(c),** showing full widths at half maxima (FWHM) of 660 nm, 580 nm, 260 nm, and 176 nm.

**MIR near-field imaging resolves the reversible photoswitching dynamics of a single lipid vesicle**

An approximately 500 nm wide vesicle was selected to map photoswitching-induced morphological changes by recording MIR images at 1603 cm$^{-1}$. (**Figs. 3a, b**). Each pair of images was acquired during intervals of approximately 2 min, with illumination periods of at least 1 min for inducing the photoswitching between each step. The time series commences with a round vesicle in the *trans*-state (**Fig. 3a, b**), followed by a blue/UV light illumination sequence to switch the photolipids multiple times between *trans* and *cis*. During the photoisomerization steps, reversible changes of the vesicle shape, directionality of deformation and size were observed.

These size expansions of the vesicle when switching from *trans* to *cis* are in good agreement with previous reports, where it was shown that vesicles undergo transformations into less symmetric shapes following *trans*-to-*cis* isomerization[19], and that the area per photolipid increases by about 20%[38] to 25 %[4] for *cis* photolipids due to a higher packing density of *trans*-azo-PC in a lipid membrane. For the *cis* isomer, the conformational change of one lipid tail (see **Fig. 1c**) would increase the chain volume, with the lipids assuming a slightly inverted wedge shape. This change in packing density is reflected in the morphological change of the vesicles, which requires a rearrangement of the lipid molecules.

To confirm that we can monitor reversible photoswitching over long times and investigate a different lipid configuration in the form of a supported membrane patch adhering to the SiN membrane, we conducted a similar photoswitching time series (**Fig. S4**). The results again highlight reversible area changes due to photoswitching over four cycles within one hour, confirming the high stability and reproducibility of the s-SNOM measurements and the good reversibility of the switching process.

Furthermore, the time-series images **(Fig. 3a, b)**, were used to extract both the vesicle's area *A* and its "circularity" as a figure of merit for the asymmetry between both states as defined in **Fig. 3c**. We delineate the vesicle's boundary by setting a threshold value of $s_2$ = 7.8 a.u.. The changes between both photoisomerization states are significant and well reproduced, amounting to an increase of 10% in area and a decrease of 8% in circularity for the *trans*-to-*cis transition*.

One possible explanation for the change in circularity is that the 50% mixture of azo-PC and DOPC used in our samples displays lower bending rigidities (which quantifies the energy needed to change the membrane curvature[39]) in the *cis* compared to the *trans*-state, which aids vesicle deformation and explains the observed change in morphology[40]. Additionally, the photostationary state (PSS) i.e. the *trans*-to-*cis*-ratio reached by the photoisomerization process, may not be quantitative, meaning that not all azo-PC lipids assume a 100% *trans*- or *cis*-conformation due to photoisomerization. The PSS is influenced by many experimental parameters, including the solvent, temperature, and the illumination conditions. For pure azo-PC vesicles in water, it was shown by small angle X-ray scattering that the fraction of *cis*-lipids in the blue adapted state (after 465 nm illumination) is still about 30%, while a considerable fraction of about 27% of the photolipids remain in a *trans*-state after UV illumination[41]. Furthermore, *trans*-lipids are prone to form H-aggregates, which may lead to phase separation and demixing of *trans*-lipids into different domains on the vesicle membrane leading to the observed expansion[42]. Our method opens the door to studying such complex photoinduced deformation effects on a multitude of other nanoscale systems such as photoswitchable metal-organic frameworks[43] and nanoparticles[44].

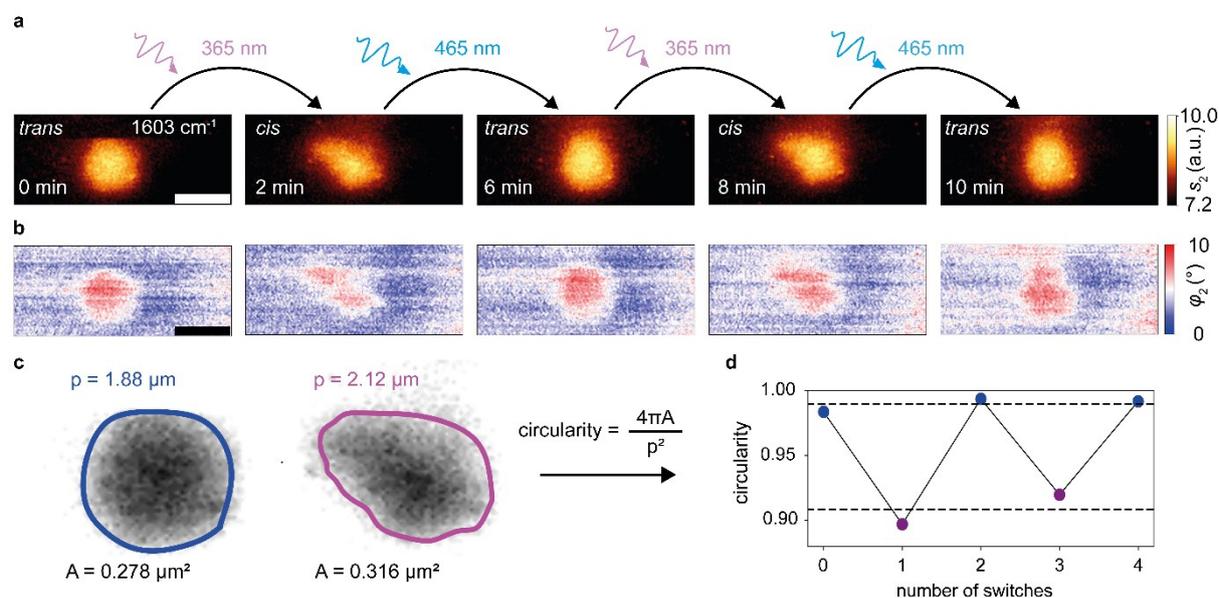

**Figure 3: Near-field imaging of the reversible photoswitching of a 500 nm diameter lipid vesicle**. **(a)** Monochromatic MIR amplitude ($s_2$) and **(b)** phase images ($\varphi_2$) at 1603 cm$^{-1}$ of a lipid vesicle in D$_2$O being reversibly photoswitched between the *trans*-state and the *cis*-state, scale bars 500 nm. **(c)** Illustrated extraction of a "circularity " value for a lipid vesicle defined as $4\pi A/p^2$, with circumference $p$ and area $A$, where the boundary criterion was $s_2$ = 7.8 a.u. **(d)** Reversible change in circularity between the *trans*- (blue) and *cis*-state (violet) of the lipid vesicle over two switching cycles.

**50 ms resolved MIR near-field-signal tracking of the photoswitching dynamics of a single lipid vesicle.**

Resolving fast photoswitching dynamics of single lipid vesicles requires nanoscopy methods that effectively track rapid structural and chemical changes. However, standard s-SNOM imaging is limited by slow mechanical scanning speeds even for comparatively low per-pixel signal integration times $t_p$, resulting in long imaging durations, for example, 50 s for a sub-µm lipid particle in water (**Figs. 3a, b**), where 150 pixel by 100 pixel were acquired using $t_p$ = 3.3 ms.

We address this inherent limitation of scanning probe methods by implementing a transient MIRnanoscopy method applicable to both aqueous and dry samples. The method continuously records the near-field s-SNOM signals $s_2$ and $\varphi_2$ at a set integration time. A specific wavelength is selected to achieve the maximum spectral response to photoswitching, and the tapping tip is placed on the center of a selected vesicle and remains stationary during the recording. This approach yields a temporal resolution of the switching dynamics at a modest SNR of about 4 down to around 50 ms (**Figs. S8, S9, S10**).

A suitable vesicle is first identified by recording an MIR image at 1603 cm$^{-1}$ (**Figs. 4a, b**). Subsequently, we confirm the vesicle's photoresponsivity by imaging the morphological distortion of the vesicle due to photoswitching. We then place the tip onto the center of the vesicle (red and black crosses in **Figs. 4a, b**) to prevent signal distortion due to movement of the vesicle. The s-SNOM signals are recorded at this defined position at 1603 cm$^{-1}$ and a time resolution of $t_P$ = 500 ms (**Fig. 4c, d**). The signal acquisition was started after illuminating the sample for over 1 min at 465 nm, followed by switching once every minute between 365 nm and 465 nm (violet and blue part of the signal trace in **Figs. 4c, d**). When the vesicle is switched to the *cis*-state, both amplitude and phase traces show clearly monotonic decreases, consistent with the recorded ATR-FTIR and nano-FTIR spectra. Likewise, both signals increase monotonically when the vesicle is switched back. Upon exposure to 365 nm light, the optical amplitude decreases by about 10%, from 1.16 to 1.07 (normalized to the average signal of D$_2$O), while the phase decreases by about 1.6°.

Notably, after an initial slow change, the signals in **Figs. 4c, d** exhibit drastically accelerated changes that last only down to 1 s in some cases and appear delayed after about 11 s in both amplitude and phase. This overall sigmoidal behavior suggests that interesting cooperative effects between the lipid molecules come into play during membrane photoswitching. Non-exponential and delayed-onset transitions have been observed for pure azo-PC vesicles by ensemble-averaged UV absorption measurements in liquid suspension[38,42]. During photoisomerization, abrupt changes of the lipid conformation along with a change of intermolecular interactions between *trans*- and *cis*-azobenzene embedded in a DOPC environment affect the isomerization dynamics[45].

Stronger lipid-lipid interactions of azo-PC are present in the *trans*-state, where the molecules form H-aggregates[42,46] and are packed more densely[38,47]. At the onset of UV illumination, the *trans-cis* ratio changes towards a *cis*-rich environment. Therefore, the photoisomerization is slower in the beginning, where a lipid conformation change is sterically hindered in the dense membrane assembly. In the extreme case of self-assembled monolayers it has been shown that steric hindrance can even prevent the *trans*-to-*cis* isomerization altogether[48]. As the membrane is shifted towards a *cis*-rich state, more space is made available, which facilitates the overall switching of the azobenzenes. In general, cis-to-trans isomerization is faster, as lipids are switched to their thermodynamically favorable state, although switching rates are strongly dependent on the illumination conditions and light intensity[38]. In the measurements presented in **Fig. 4**, however, it seems that both isomerization rates are almost similar. Furthermore, a delay was also observed for the onset of *cis*-to-*trans* isomerization, which likely originates from membrane reorganization taking place as the photolipids reduce their lipid footprint accompanied by an overall reduction of the bilayer fluidity[40] and vesicle deformation (**Fig. 3**). The abrupt change of the isomerization curve observed in both directions (**Fig. 4c, d**) has not been observed in absorption measurements of vesicle solutions. However, the finding is consistent with the switching dynamics observed by time-resolved monitoring of supported photolipid bilayer isomerization by single particle plasmonic sensing[49]. This supports our understanding that nanoscopy on a single vesicle level, where real-time information is obtained from a localized membrane area under homogeneous illumination, provides further insight to otherwise hidden details of photoisomerization dynamics of photolipids in a bilayer assembly.

Remarkably, this back-and-forth switching could be continuously monitored over an 8 min duration, encompassing four complete switching cycles. After recording the time trace, we recorded another MIR image (**Fig. 4e, f**) to verify that the vesicle has not shifted in position relative to the tip, thus eliminating the possibility of signal fluctuations due to vesicle displacement during the switching process.

For comparison, we recorded a second transient signal trace at 1730 cm$^{-1}$ (**Fig. S5**), probing the carbonyl resonance that should remain unaffected by the photoswitching-induced molecular changes observed in **Fig. S1**. We measured only relatively small amplitude and phase signal changes, verifying that at 1603 cm$^{-1}$ we did probe the molecular switching. We attributed the recorded decrease in amplitude and phase from the *trans*- to the *cis*-state to a reduced lipid density in the *cis*-state[42]. Moreover, we recorded a signal trace on $D_2O$ next to the vesicle (**Fig. S6**) and on a clean silicon surface (**Fig. S7**). Both traces exhibited stable amplitude and phase signals, confirming that the measured switching signals are not due to mechanical or optical artifacts induced by the switching light.

To explore the temporal resolution limits of our in-situ tracking of dynamic processes, we repeated photoswitching experiments at different integration times $t_p$ ranging from 30 ms to 500 ms (**Figs. S8, S9, S10**). The signal-to-noise characteristics allowed to detect a photoswitching signature at a temporal resolution as short as 30 ms with a statistical significance limit of 3.72 $\sigma$ (see **Fig. S10 and Fig. S8 and S9** for exact values). Notably, by increasing the integration time to 100 ms (**Fig. S8 and S9**) the SNR jumps to 5.5 $\sigma$, showing the great potential for the technique to also be applied to dried samples such as for example the investigation of photoactive proteins at the single-molecule level[35]. Since these experiments do not need to be conducted in aqueous environment, even higher SNR as shown here could be expected.

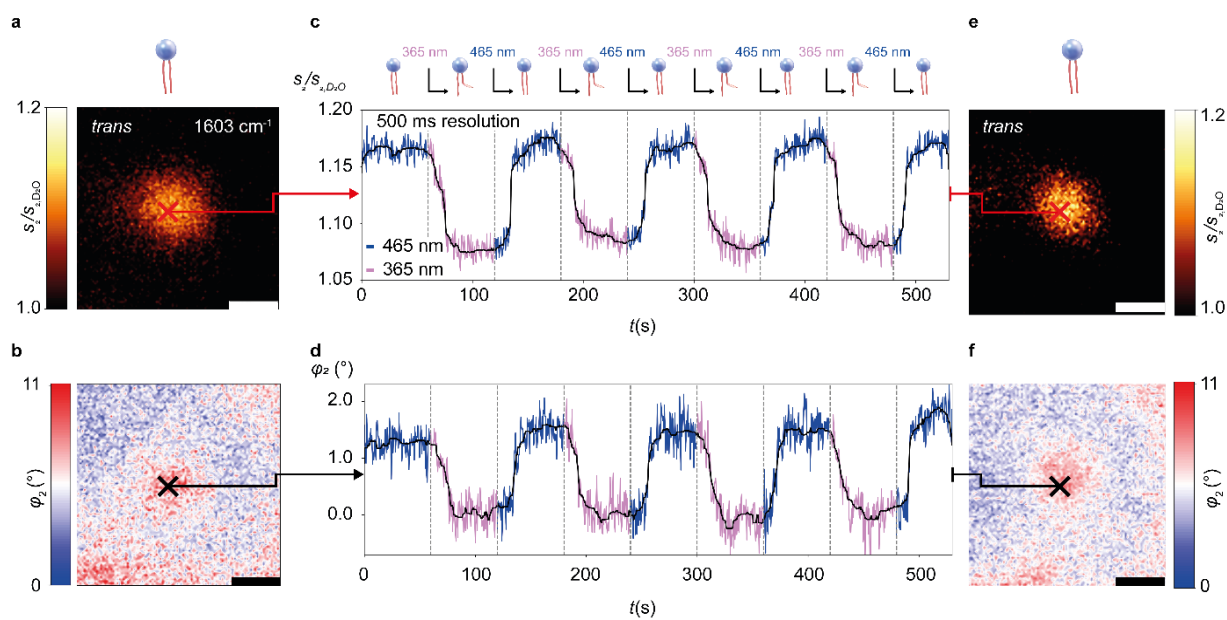

**Figure 4: Resolving the photoswitching dynamics of a single nanoscale lipid vesicle by millisecond MIR near-field signal traces. (a)** Optical near-field amplitude image normalized to the surrounding $D_2O$ ($s_2/s_{2,\,D_2O}$) and **(b)** near-field phase image ($\varphi_2$) recorded at 1603 cm$^{-1}$ with 2 min acquisition time, directly before starting the MIR signal trace acquisition at the position of the cross. The recorded amplitude ($s_2/s_{2,D2O}$) in **(c)** together with the phase ($\varphi_2$) in **(d)** reveal the photoswitching dynamics at a temporal resolution of 500 ms. The symbols above **(c)** in combination with the dashed vertical lines indicate when the illumination wavelength was switched to the value written above. The blue and violet coloring of the signal traces mark the switching light's wavelength at each point. The black curves are moving averages of the measured points. **(e)** Optical amplitude and **(f)** phase images taken directly after acquiring the near-field traces show that the vesicle remained stable in both position and signal strength, even after seven consecutive switching transitions.

## 3. Discussion

By combining near-field microscopy with ultrathin membranes and fast signal acquisition methods, we have demonstrated the label-free imaging and spectroscopic detection of actively triggered photoswitching processes in lipid nanovesicles as small as 176 nm in their native aqueous environment. Specifically, we utilized our method to resolve reversible photoinduced shape changes in the vesicles over multiple switching cycles. This capability enables future studies of shape effects in photorelease processes for a wide range of nanocarriers. Furthermore, leveraging the inherent spectral and therefore chemical sensitivity of MIR nanoscopy, even at the single lipid vesicle level, we could differentiate two main photoisomeric states of the photoswitchable vesicles based on nano-FTIR spectra. The chemical specificity of this method holds great potential for investigating other complex lipid systems on the nanoscale, such as the chemical composition of domains in lipid vesicles[42] or dynamics of lipid nanoparticles loaded with therapeutic compounds like mRNA.

Additional information about the geometry of the vesicles was obtained by comparing measured phase profiles of a single vesicle with an analytical model for a nondeformable sphere, suggesting a distinct flattening of the vesicles when adhering to the membrane. The vesicles were also identifiable in the correlative mechanical images, where we attribute the upward displacement of the membrane to van der Waals forces exerted by the vesicle.

Importantly, we implemented a transient MIR nanoscopy technique to extract the switching dynamics of a single lipid vesicle with millisecond time resolution. It is significant to note that we conducted the experiments on vesicular systems in liquid, which exhibit rather weak back-scattering. One way to increase the scattering signals and with it the time resolution by as much as an order of magnitude would be the use of broader tips (albeit leading to reduced spatial resolution).[50] With other, more strongly scattering objects, the time resolution would potentially improve to the µs range.

This could be particularly interesting for electrically modulated solid-state materials[51], and would require higher phase modulation frequencies offered by photo-elastic or acousto-optic modulators[52,53]. Moreover, improvements in MIR laser stability and power, in addition to optimizing collection optics, could increase the SNR and lead to shorter acquisition times.

In applications requiring super-resolution imaging, using finer tips in our s-SNOM measurements (ideally down to about 30 nm) could significantly improve both the optical and the mechanical spatial resolution. This would also reduce the near-field probing depth to 30 nm, which would be ideally suited to image the adhering portion of a vesicle's membrane with improved contrast. Additionally, this approach could be key to detecting inhomogeneities like lipid rafts and other nanostructures[50].

Note that in principle, our method is not limited to either relatively slow imaging (**Figs. 3a, b**) or fast one-pixel tracing (**Figs. 4c, d**), but could combine both for a truly spatio-temporal assessment in real time. When extended to perform rapidly repeated, short line scans across a vesicle's edge, our method could capture abrupt spectroscopic changes inside and simultaneously morphological expansions to the outside of a vesicle in a correlative manner[8].

As a future prospect, the stable membrane-based in situ s-SNOM technique demonstrated in this work can be integrated with state-of-the-art microfluidics and environmental controls. This integration holds great promise for the nanoscale investigation of dynamic biochemical processes actively triggered by changes in temperature, pH, osmolality, or the injection of chemical compounds. Additionally, the method is compatible with correlated measurements such as fluorescence and Raman imaging.

Finally, we believe that the accuracy and versatility of our method opens up unprecedented avenues for future studies of even more complex lipid-associated phenomena and a multitude of other highly relevant systems, ranging from dynamic metal-organic-frameworks in chemistry[43,54], the assembly of Alzheimer-associated peptides in medicine[55], to the electronic modulation of 2D materials in physics[51].

## 4. Materials and Methods

**Near-field optical microscopy setup for in-situ photoswitching**

The measurements were performed using a commercially available s-SNOM in reflection mode (neaSCOPE, *attocube systems*, Haar, Germany), which records correlated AFM-topographic and near-field optical measurements. All experiments were conducted with PtIr-coated AFM tips with a tip radius of 60 nm used as a scattering probe (nanoFTIR-tips, *attocube systems*, Haar, Germany) operated in tapping mode with a tapping amplitude of 80 nm a tapping frequency $\Omega$ of approximately 250 kHz at about 85% of the resonance frequency. The optical signal from the tip is attained by focusing laser light onto the tip and collimating the backscattering of the scanning probe via a parabolic mirror. Subsequently, the backscattered light is detected by a nitrogen-cooled mercury cadmium telluride (MCT) detector (IR-20-00103, *Infrared Associates Inc.*, Stuart, USA) and demodulated at harmonics $n\Omega$ of the tapping frequency $\Omega$ to separate the near-field signal from unwanted far-field signals. The near-field optical amplitude and phase images and the near-field signal traces at a defined tip position are recorded with a pseudo-heterodyne interferometric detection scheme[56], using an optical parametric oscillator laser (OPO) with a 1050 nm pump laser and an ultra-broadly tunable MIR output ranging from $\lambda = 1.4$ μm to $\lambda = 16$ μm attained by difference frequency generation (DFG) in a non-linear crystal (Alpha with MIR extension, *Stuttgart Instruments*, Stuttgart, Germany). The resonance-specific IR images are achieved by using a grating monochromator to limit the spectral bandwidth to 10 cm$^{-1}$ and tuning the laser to the desired wavelength of the molecular absorption. The output power of the laser is attenuated by a wire grating (*Lasnix*, Berg, Germany) to about 3 mW before reaching the beamsplitter. The pseudo-heterodyne detection allows for the decoupling of the demodulated complex-valued scattering coefficient $\sigma_n(\omega)$ into optical amplitude $s_n(\omega)$ and phase $\varphi_n(\omega)$ for a chosen wavelength[56]. The detected optical amplitude relates to the reflectivity and the optical phase to the IR absorption of the sample under the tip position[57].

The nano-FTIR spectra and white light images were acquired with a broadband IR laser source based on difference frequency generation (DFG) (FFdichro_midIR_NEA 31002, *Toptica Photonics*, Martinsried, Germany)[58]. The output power was about 400 µW, and the covered spectral range between 1200 and 2000 cm$^{-1}$ (7.6–5 µm), achieving maximum power at 1666 cm$^{-1}$ (output mode D). The backscattering from the tip was analyzed by an asymmetric Michelson interferometer with the tip and sample in one and a moveable reference mirror in the other arm of the interferometer. The detected signal was demodulated at harmonics n of the tapping frequency nΩ and Fourier-transformed to obtain spectra of the optical amplitude $s_n$ and phase $\varphi_n$, which are referenced to spectra recorded on a clean silicon surface to eliminate atmospheric and instrumental artifacts in the measured spectra[34]. The near-field phase spectra are best compared to far-field ATR-FTIR spectra since ATR-FTIR is a technique that uses the penetration of evanescent waves through an interface to probe the sample, and thus the recorded ATR-FTIR spectra closely resemble the tip-enhanced near-field phase spectra. The white light images are obtained by placing the reference mirror of the asymmetric Michelson interferometer at the position of maximal constructive interference. The resulting image represents the averaged backscattering of the sample below the tip over the broad laser output spectrum and is not specific to a particular IR resonance. The white-light images are used to identify the vesicles in this study, and the tip is placed at the proper position to record the nano-FTIR spectra subsequently.

The photoswitching was performed by alternating between two diodes emitting at 365 nm and 465 nm wavelength, respectively (*Prizmatix*, Holon, Israel). The illumination power was set to around 30 mW, and the beam was concentrated by a section of the parabolic mirror as sketched in **Fig. 1b**, ensuring that the SiN membrane around the tip was homogeneously irradiated over an area of around 0.5 x 0.5 mm$^2$.

**Liquid s-SNOM sample preparation**

Liquid s-SNOM samples were prepared by first pretreating the cavity side of a 10 nm thin and 250x250 µm wide SiN membrane pretensioned on a commercial silicon chip (NX5025Z, *Norcada*, Edmonton, Canda) with a 30 min UV-Ozone cleaning procedure to increase the wettability of the membrane, then mounting the chip on a custom-made membrane holder and drop-casting 15 µL of the sample into the cavity side of the membrane.[8] After several minutes of incubation, the sample holder was sealed to prevent the evaporation of water, and placed in the s-SNOM system.

**ATR-FTIR Spectroscopy**

ATR-FTIR spectra were recorded with a commercial spectrometer (Frontier™ FT-IR spectrometer, *PerkinElmer*, Waltham, USA) with a proprietary KBr beamsplitter and a diamond ATR crystal. The spectra were recorded with a wavenumber range from 650 cm$^{-1}$ to 4000 cm$^{-1}$ with a spectral resolution of 4 cm$^{-1}$ and averaging over 60 scans.

**Finite dipole model for s-SNOM**

Analytical calculations of near-field response were conducted using Python. To predict the optical response of a multi-layered sample, the finite dipole model of near-field interaction of a metal ellipsoid with a nearby material was extended to multiple layers via the transfer matrix method[8,59]. Results are shown in Fig. S3, using the modeling parameters $a$ = 80 nm (tapping amplitude), $r$ = 60 nm (tip radius), $L$ = 300 nm (length of spheroid), $g$ = 0.6 (charge induced in the tip).

**Preparation of Liposomes**

The photolipid azo-PC was synthesized according to a previous protocol[19]. Vesicles were prepared by electroformation using a vesicle prep pro device (*Nanion Technologies*, Munich, Germany). Two conductive glass substrates (coated by Indium tin Oxide, ITO) were separated by an O-ring, forming a sandwich-constructed chamber. Lipids were dissolved in chloroform at a concentration of 10 mM. 20 µL of the lipid solution (50% DOPC (*Merck*, Darmstadt, Germany) and 50% azo-PC) were added on the conductive side of ITO substrate within the O-ring. After evaporation of the chloroform, the O-ring chamber was filled with 250 µL of sucrose solution at 300 mM concentration. Sucrose solutions were prepared with $H_2O$ or $D_2O$. Electroformation was conducted by applying an electric field (5 Hz, 3 V) at 37 °C for 120 min.

**Data availability statement**

The main data supporting the findings of this study are available within the article and its Supplementary Information files.

**Acknowledgment**

The authors would like to thank J. Rädler and B. Nickel for valuable discussions. This project was funded by the Deutsche Forschungsgemeinschaft (DFG, German Research Foundation) under grant number TI 1063/1 (Emmy Noether Program) and the Center for NanoScience (CeNS). Funded by the European Union (ERC, METANEXT, 101078018 and EIC, NEHO, 101046329). Views and opinions expressed are however those of the author(s) only and do not necessarily reflect those of the European Union, the European Research Council Executive Agency, or the European Innovation Council and SMEs Executive Agency (EISMEA). Neither the European Union nor the granting authority can be held responsible for them. TL was supported by the Deutsche Forschungsgemeinschaft (DFG) through the Collaborative Research Center SFB1032 (Project No. 201269156, Project A8). JZ is supported by the China Scholarship Council. SAM additionally acknowledges the Lee-Lucas Chair in Physics.

**Author contributions**

T.G., F.K., T.L and A.T. conceived the idea and planned the project. T.G., E.B and K.K. performed the near-field and far-field measurements. J.Z. and D.T. provided and prepared the photoswitchable lipid sample. E.B. performed analytical simulations. T.G., E.B., F.K, J.Z., T.L, A.T helped with analyzing the experimental data. T.G., F.K, and A.T. wrote the manuscript with input from all the authors. A.T., T.L., F.K., and S.A.M. managed and supervised various aspects of the project.

**Conflict of interest**

F. Keilmann is a scientific advisor to attocube systems AG, manufacturer of the s-SNOM used in this study. T. Gölz obtained financial support for his PhD thesis from attocube systems AG. K. Kaltenecker is employee of attocube systems AG.

**Supporting information:**

**References**
1. DiFrancesco, M. L. *et al.* Neuronal firing modulation by a membrane-targeted photoswitch. *Nature nanotechnology* **15,** 296–306; 10.1038/s41565-019-0632-6 (2020).

2. Chander, N. *et al.* Optimized Photoactivatable Lipid Nanoparticles Enable Red Light Triggered Drug Release. *Small (Weinheim an der Bergstrasse, Germany)* **17,** e2008198; 10.1002/smll.202008198 (2021).

3. Chen, H., Zhang, W., Zhu, G., Xie, J. & Chen, X. Rethinking cancer nanotheranostics. *Nat Rev Mater* **2**; 10.1038/natrevmats.2017.24 (2017).

4. Aleksanyan, M. *et al.* Photomanipulation of Minimal Synthetic Cells: Area Increase, Softening, and Interleaflet Coupling of Membrane Models Doped with Azobenzene-Lipid Photoswitches. *Advanced Science* **10**; 10.1002/advs.202304336 (2023).

5. Broichhagen, J., Frank, J. A. & Trauner, D. A roadmap to success in photopharmacology. *Accounts of chemical research* **48,** 1947–1960; 10.1021/acs.accounts.5b00129 (2015).

6. Keilmann, F. & Hillenbrand, R. Near-field microscopy by elastic light scattering from a tip. *Philosophical transactions. Series A, Mathematical, physical, and engineering sciences* **362,** 787–805; 10.1098/rsta.2003.1347 (2004).

7. Chen, X. *et al.* Modern Scattering-Type Scanning Near-Field Optical Microscopy for Advanced Material Research. *Advanced materials (Deerfield Beach, Fla.)* **31,** e1804774; 10.1002/adma.201804774 (2019).

8. Kaltenecker, K. J., Gölz, T., Bau, E. & Keilmann, F. Infrared-spectroscopic, dynamic near-field microscopy of living cells and nanoparticles in water. *Scientific reports* **11,** 21860; 10.1038/s41598-021-01425-w (2021).

9. Qu, D.-H., Wang, Q.-C., Zhang, Q.-W., Ma, X. & Tian, H. Photoresponsive Host-Guest Functional Systems. *Chemical reviews* **115,** 7543–7588; 10.1021/cr5006342 (2015).

10. He, X., Larson, J. M., Bechtel, H. A. & Kostecki, R. In situ infrared nanospectroscopy of the local processes at the Li/polymer electrolyte interface. *Nature communications* **13,** 1398; 10.1038/s41467-022-29103-z (2022).

11. Karst, J. *et al.* Watching in situ the hydrogen diffusion dynamics in magnesium on the nanoscale. *Science advances* **6,** eaaz0566; 10.1126/sciadv.aaz0566 (2020).

12. Krüger, A., Bürkle, A., Hauser, K. & Mangerich, A. Real-time monitoring of PARP1-dependent PARylation by ATR-FTIR spectroscopy. *Nature communications* **11,** 2174; 10.1038/s41467-020-15858-w (2020).

13. Güldenhaupt, J. *et al.* Ligand-Induced Conformational Changes in HSP90 Monitored Time Resolved and Label Free-Towards a Conformational Activity Screening for Drug Discovery. *Angewandte Chemie (International ed. in English)* **57,** 9955–9960; 10.1002/anie.201802603 (2018).

14. Allen, T. M. & Cullis, P. R. Liposomal drug delivery systems: from concept to clinical applications. *Advanced drug delivery reviews* **65,** 36–48; 10.1016/j.addr.2012.09.037 (2013).

15. Nsairat, H. *et al.* Liposomes: structure, composition, types, and clinical applications. *Heliyon* **8,** e09394; 10.1016/j.heliyon.2022.e09394 (2022).

16. Lamichhane, N. *et al.* Liposomes: Clinical Applications and Potential for Image-Guided Drug Delivery. *Molecules (Basel, Switzerland)* **23**; 10.3390/molecules23020288 (2018).

17. Swetha, K. *et al.* Recent Advances in the Lipid Nanoparticle-Mediated Delivery of mRNA Vaccines. *Vaccines* **11**; 10.3390/vaccines11030658 (2023).

18. Sercombe, L. *et al.* Advances and Challenges of Liposome Assisted Drug Delivery. *Frontiers in pharmacology* **6,** 286; 10.3389/fphar.2015.00286 (2015).

19. Pernpeintner, C. *et al.* Light-Controlled Membrane Mechanics and Shape Transitions of Photoswitchable Lipid Vesicles. *Langmuir : the ACS journal of surfaces and colloids* **33,** 4083–4089; 10.1021/acs.langmuir.7b01020 (2017).

20. Pritzl, S. D. *et al.* Optical Membrane Control with Red Light Enabled by Red-Shifted Photolipids. *Langmuir : the ACS journal of surfaces and colloids* **38,** 385–393; 10.1021/acs.langmuir.1c02745 (2022).

21. Stuart, B. H. *Infrared spectroscopy. Fundamentals and applications* (Wiley, Chichester, 2008).

22. Baserga, F. *et al.* Membrane Protein Activity Induces Specific Molecular Changes in Nanodiscs Monitored by FTIR Difference Spectroscopy. *Frontiers in molecular biosciences* **9,** 915328; 10.3389/fmolb.2022.915328 (2022).

23. Crea, F. *et al.* Photoactivation of a Mechanosensitive Channel. *Frontiers in molecular biosciences* **9,** 905306; 10.3389/fmolb.2022.905306 (2022).

24. Amenabar, I. *et al.* Structural analysis and mapping of individual protein complexes by infrared nanospectroscopy. *Nature communications* **4,** 2890; 10.1038/ncomms3890 (2013).

25. Cernescu, A. *et al.* Label-Free Infrared Spectroscopy and Imaging of Single Phospholipid Bilayers with Nanoscale Resolution. *Analytical chemistry* **90,** 10179–10186; 10.1021/acs.analchem.8b00485 (2018).

26. Autret, M., Le Plouzennec, M., Moinet, C. & Simonneaux, G. Intramolecular fluorescence quenching in azobenzene-substituted porphyrins. *J. Chem. Soc., Chem. Commun.,* 1169; 10.1039/C39940001169 (1994).

27. Hartrampf, N. *et al.* Structural diversity of photoswitchable sphingolipids for optodynamic control of lipid microdomains. *Biophysical journal* **122,** 2325–2341; 10.1016/j.bpj.2023.02.029 (2023).

28. Stollmann, A. *et al.* Molecular fingerprinting of biological nanoparticles with a label-free optofluidic platform. *ArXiv* (2023).

29. Bello, V. *et al.* Transmission electron microscopy of lipid vesicles for drug delivery: comparison between positive and negative staining. *Microscopy and microanalysis : the official journal of Microscopy Society of America, Microbeam Analysis Society, Microscopical Society of Canada* **16,** 456–461; 10.1017/S1431927610093645 (2010).

30. Rodrigo, D. *et al.* Resolving molecule-specific information in dynamic lipid membrane processes with multi-resonant infrared metasurfaces. *Nature communications* **9,** 2160; 10.1038/s41467-018-04594-x (2018).

31. Wittenberg, N. J. *et al.* High-affinity binding of remyelinating natural autoantibodies to myelin-mimicking lipid bilayers revealed by nanohole surface plasmon resonance. *Analytical chemistry* **84,** 6031–6039; 10.1021/ac300819a (2012).

32. Limaj, O. *et al.* Infrared Plasmonic Biosensor for Real-Time and Label-Free Monitoring of Lipid Membranes. *Nano letters* **16,** 1502–1508; 10.1021/acs.nanolett.5b05316 (2016).

33. Quidant, R. Plasmon Nano-Optics: Designing Novel Nano-Tools for Biology and Medicine. In *Plasmonics,* edited by S. Enoch & N. Bonod (Springer Berlin Heidelberg, Berlin, Heidelberg, 2012), Vol. 167, pp. 201–222.

34. Amarie, S. & Keilmann, F. Broadband-infrared assessment of phonon resonance in scattering-type near-field microscopy. *Phys. Rev. B* **83**; 10.1103/PhysRevB.83.045404 (2011).

35. Lorenz-Fonfria, V. A. Infrared Difference Spectroscopy of Proteins: From Bands to Bonds. *Chemical reviews* **120,** 3466–3576; 10.1021/acs.chemrev.9b00449 (2020).

36. Mester, L., Govyadinov, A. A., Chen, S., Goikoetxea, M. & Hillenbrand, R. Subsurface chemical nanoidentification by nano-FTIR spectroscopy. *Nature communications* **11,** 3359; 10.1038/s41467-020-17034-6 (2020).

37. Baù, E., Gölz, T., Benoit, M., Tittl, A. & Keilmann, F. Nanoscale mechanical manipulation of ultrathin SiN membranes enabling infrared near-field microscopy of liquid-immersed samples, Preprint at https://doi.org/10.48550/arXiv.2404.01770 (2024).

38. Pritzl, S. D. et al. Photolipid Bilayer Permeability is Controlled by Transient Pore Formation. *Langmuir : the ACS journal of surfaces and colloids* **36,** 13509–13515; 10.1021/acs.langmuir.0c02229 (2020).

39. Faizi, H. A., Frey, S. L., Steinkühler, J., Dimova, R. & Vlahovska, P. M. Bending rigidity of charged lipid bilayer membranes. *Soft matter* **15,** 6006–6013; 10.1039/C9SM00772E (2019).

40. Urban, P. et al. A Lipid Photoswitch Controls Fluidity in Supported Bilayer Membranes. *Langmuir : the ACS journal of surfaces and colloids* **36,** 2629–2634; 10.1021/acs.langmuir.9b02942 (2020).

41. Ober, M. F. et al. SAXS measurements of azobenzene lipid vesicles reveal buffer-dependent photoswitching and quantitative Z→E isomerisation by X-rays. *Nanophotonics* **11,** 2361–2368; 10.1515/nanoph-2022-0053 (2022).

42. Urban, P. et al. Light-Controlled Lipid Interaction and Membrane Organization in Photolipid Bilayer Vesicles. *Langmuir : the ACS journal of surfaces and colloids* **34,** 13368–13374; 10.1021/acs.langmuir.8b03241 (2018).

43. Rice, A. M. et al. Photophysics Modulation in Photoswitchable Metal-Organic Frameworks. *Chemical reviews* **120,** 8790–8813; 10.1021/acs.chemrev.9b00350 (2020).

44. Tong, R., Hemmati, H. D., Langer, R. & Kohane, D. S. Photoswitchable nanoparticles for triggered tissue penetration and drug delivery. *Journal of the American Chemical Society* **134,** 8848–8855; 10.1021/ja211888a (2012).

45. Osella, S., Granucci, G., Persico, M. & Knippenberg, S. Dual photoisomerization mechanism of azobenzene embedded in a lipid membrane. *Journal of materials chemistry. B* **11,** 2518–2529; 10.1039/D2TB02767D (2023).

46. Kuiper, J. M. & Engberts, J. B. F. N. H-aggregation of azobenzene-substituted amphiphiles in vesicular membranes. *Langmuir : the ACS journal of surfaces and colloids* **20,** 1152–1160; 10.1021/la0358724 (2004).

47. Kuiper, J. M., Stuart, M. C. A. & Engberts, J. B. F. N. Photochemically induced disturbance of the alkyl chain packing in vesicular membranes. *Langmuir : the ACS journal of surfaces and colloids* **24,** 426–432; 10.1021/la702892m (2008).

48. Valley, D. T., Onstott, M., Malyk, S. & Benderskii, A. V. Steric hindrance of photoswitching in self-assembled monolayers of azobenzene and alkane thiols. *Langmuir : the ACS journal of surfaces and colloids* **29,** 11623–11631; 10.1021/la402144g (2013).

49. Zhang, J. *et al.* Label-Free Time-Resolved Monitoring of Photolipid Bilayer Isomerization by Plasmonic Sensing. *Advanced Optical Materials* **12**; 10.1002/adom.202302266 (2024).

50. Maissen, C., Chen, S., Nikulina, E., Govyadinov, A. & Hillenbrand, R. Probes for Ultrasensitive THz Nanoscopy. *ACS Photonics* **6,** 1279–1288; 10.1021/acsphotonics.9b00324 (2019).

51. Fei, Z. *et al.* Gate-tuning of graphene plasmons revealed by infrared nano-imaging. *Nature* **487,** 82–85; 10.1038/nature11253 (2012).

52. Hillenbrand, R., Knoll, B. & Keilmann, F. Pure optical contrast in scattering-type scanning near-field microscopy. *Journal of microscopy* **202,** 77–83; 10.1046/j.1365-2818.2001.00794.x (2001).

53. Pfitzner, E. Surface and Tip-Enhanced Infrared Spectroscopy in Life Science, 2019.

54. Möslein, A. F., Gutiérrez, M., Cohen, B. & Tan, J.-C. Near-Field Infrared Nanospectroscopy Reveals Guest Confinement in Metal-Organic Framework Single Crystals. *Nano letters* **20,** 7446–7454; 10.1021/acs.nanolett.0c02839 (2020).

55. Perálvarez-Marín, A., Barth, A. & Gräslund, A. Time-resolved infrared spectroscopy of pH-induced aggregation of the Alzheimer Abeta(1-28) peptide. *Journal of molecular biology* **379,** 589–596; 10.1016/j.jmb.2008.04.014 (2008).

56. Vicentini, E. *et al.* Pseudoheterodyne interferometry for multicolor near-field imaging. *Optics express* **31,** 22308–22322; 10.1364/OE.492213 (2023).

57. Huth, F. *et al.* Nano-FTIR absorption spectroscopy of molecular fingerprints at 20 nm spatial resolution. *Nano letters* **12,** 3973–3978; 10.1021/nl301159v (2012).

58. Amarie, S., Ganz, T. & Keilmann, F. Mid-infrared near-field spectroscopy. *Optics express* **17,** 21794–21801; 10.1364/OE.17.021794 (2009).

59. Hauer, B., Engelhardt, A. P. & Taubner, T. Quasi-analytical model for scattering infrared near-field microscopy on layered systems. *Optics express* **20,** 13173–13188; 10.1364/OE.20.013173 (2012).

# Supplementary information

# Transient infrared nanoscopy resolves the millisecond photoswitching dynamics of single lipid vesicles in water

*T. Gölz[1], E. Baù[1], J. Zhang[2], K. Kaltenecker[1,3], D. Trauner[4], S. A. Maier[5,6], F. Keilmann[1*], T. Lohmueller[2*], A. Tittl[1*]*

1. Chair in Hybrid Nanosystems, Nano-Institute Munich, Department of Physics, Ludwig-Maximilians-Universität München, 80539 Munich, Germany

2. Chair for Photonics and Optoelectronics, Nano-Institute Munich, Department of Physics, Ludwig-Maximilians-Universität München, 80539 Munich, Germany

3. Attocube Systems AG, 85540 Haar, Germany

4. Department of Chemistry, University of Pennsylvania, Philadelphia, Pennsylvania 19104-6323, United States

5. School of Physics and Astronomy, Monash University, Clayton, Victoria 3800, AUS

6. Department of Physics, Imperial College London, London SW7 2AZ, UK

Email: fritz.keilmann@lmu.de, t.lohmueller@lmu.de, andreas.tittl@physik.uni-muenchen.de

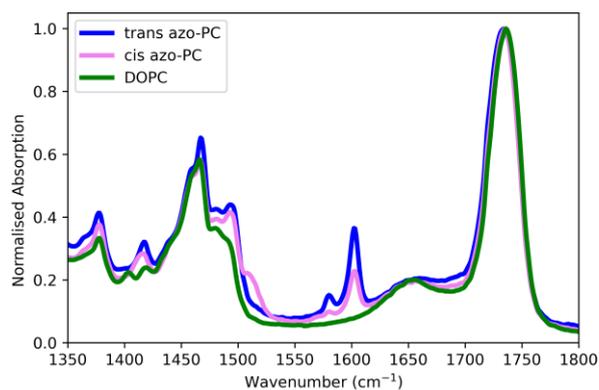

**Figure S1:** Normalised ATR-FTIR spectra of DOPC (green) and azo-PC in the *trans*-(blue) and *cis*-state (violet) measured on dried samples. The data of each spectrum are normalised to the corresponding carbonyl resonance at around 1735 cm$^{-1}$ to compare the peak intensities between the different spectra. The photoswitching between the *cis*/*trans*-state was performed by illuminating the dried sample on the ATR-crystal with 365 nm (*trans* to *cis*) and 465 nm light (*cis* to *trans*).

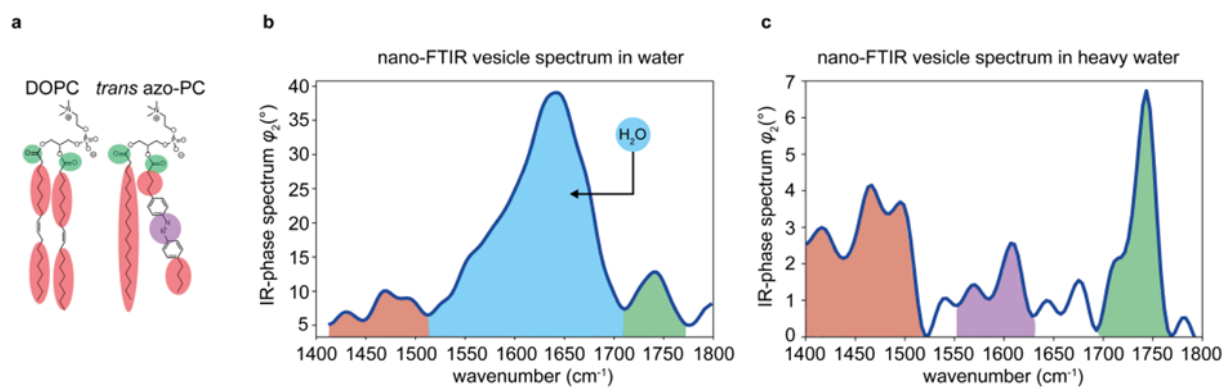

**Figure S2: (a)** Molecular sketches of DOPC and *trans*-azo-PC composing the lipid vesicles with bonds highlighted in specific colours to assign them to the resonances in the nano-FTIR phase spectra (**b, c**). Experimentally determined nano-FTIR phase spectra ($\varphi_2$) of a lipid vesicle suspended in H$_2$O **(b)** and D$_2$O **(c)**.

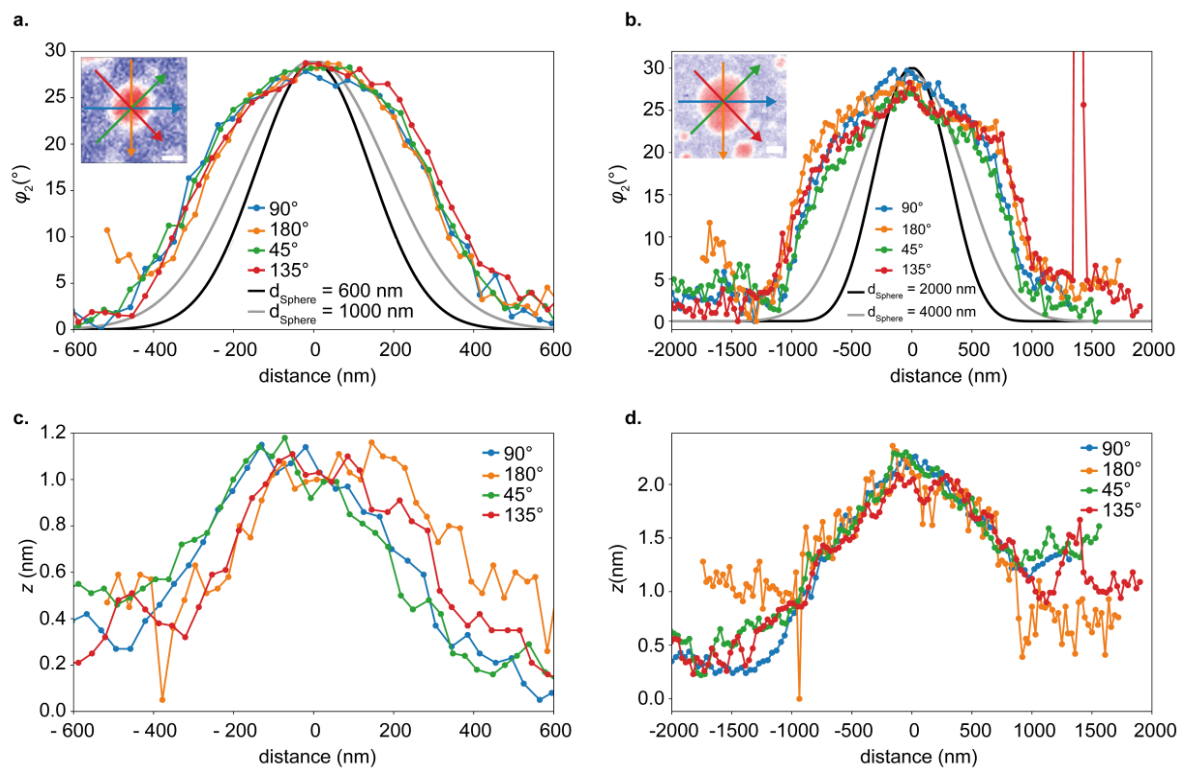

**Figure S3**: **(a, b)** Profiles of the green-boxed vesicle and the largest vesicle (next to the brown box) in the phase image shown in **Fig. 2b,** extracted along the arrows indicated in the inset (scale bars 300 nm and 1 μm, respectively). **(c, d)** Correlative topography profiles from the simultaneously measured topographic images (not shown). The theoretical curves in **(a, b)** result from analytically predicting the phase profiles of differently sized, undeformed spheres ($d_{Sphere}$= 600nm, 1000 nm, 2000 nm and 4000 nm, material PMMA, tip radius = 60 nm and tapping amplitude a = 80 nm) which are assumed to adhere at one point on the lower surface of a 10 nm SiN membrane (for details see **Materials and Methods** and previous literature[1,2]).

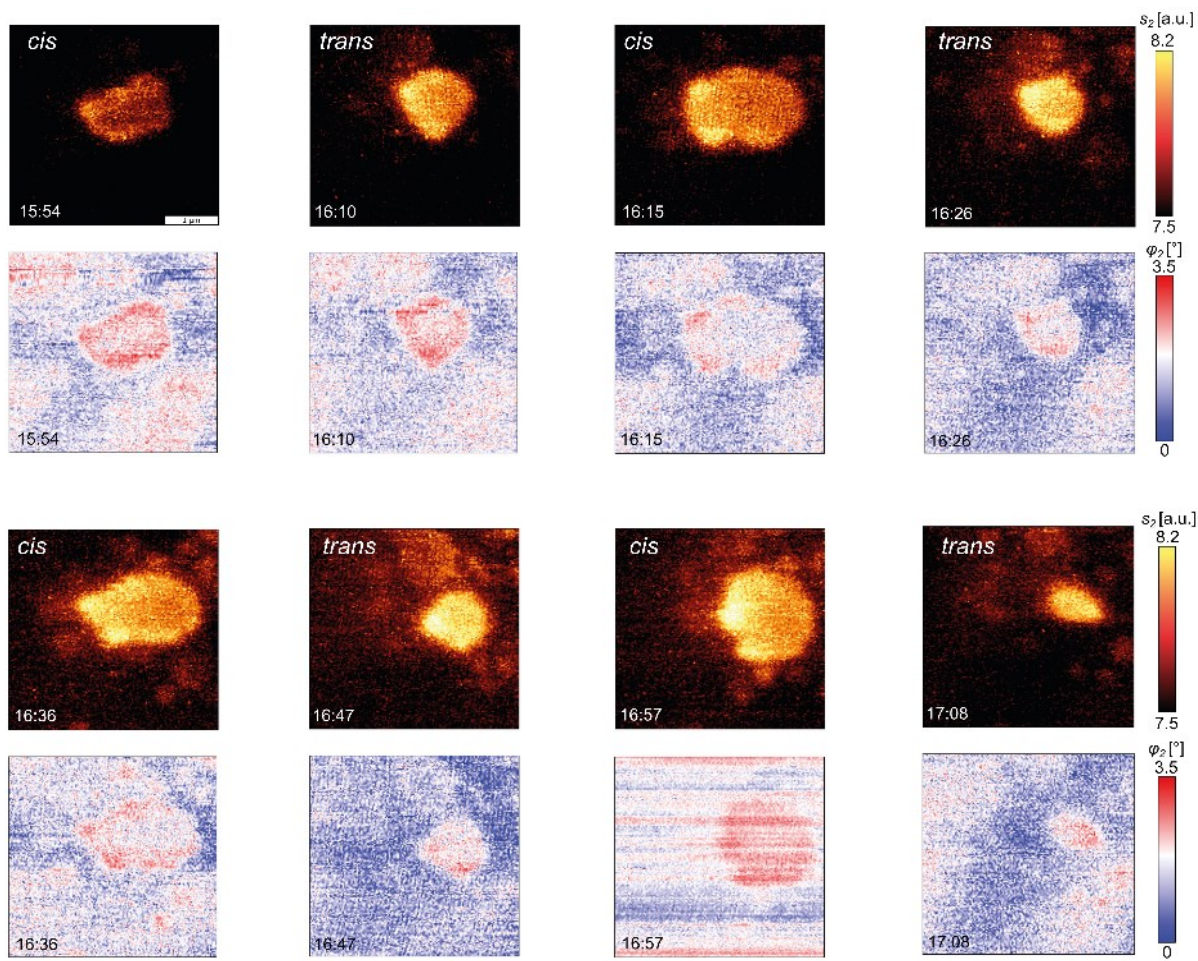

**Figure S4:** Optical amplitude ($s_2$) and phase ($\varphi_2$) image series of a lipid patch being reversible photoswitched between the *cis* and *trans*-state showing a reversible expansion (*cis*-state) and contraction (*trans*-state) over a 1 h measurement time with a scale bar of 1 μm for all images.

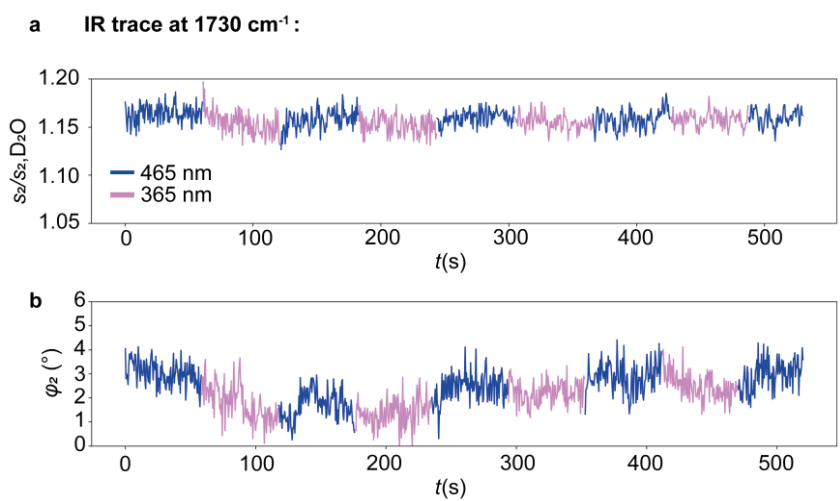

**Figure S5:** Near-field signal trace of the optical amplitude normalised to the D$_2$O signal ($s_2/s_{2,D_2O}$) **(a)** and of the optical phase ($\varphi_2$) **(b)** recorded at 1730 cm$^{-1}$ on the vesicle shown in **Figs. 4a, b, e** and **f**. The blue and violet colouring of the IR signal trace specifies the wavelength of the switching light of 465 nm and 365 nm, respectively.

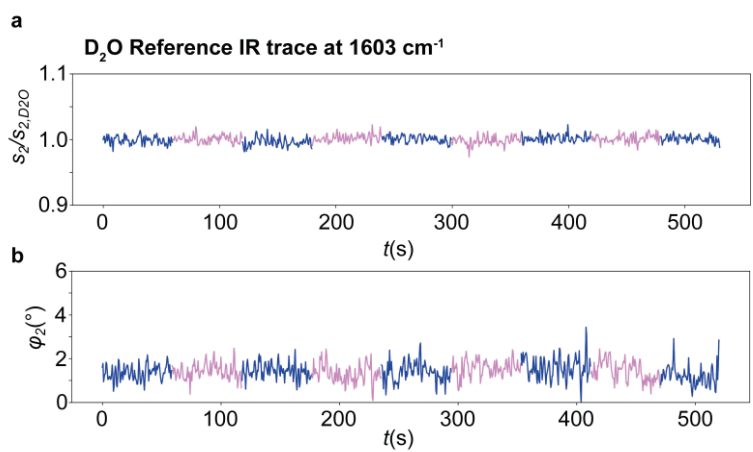

**Figure S6:** Near-field optical amplitude normalised to the D$_2$O signal ($s_2/s_{2,D2O}$) **(a)** and phase ($\varphi_2$) **(b)**, recorded at 1603 cm$^{-1}$ on D$_2$O next to the vesicle shown in **Figs. 4 a** and **b**. The blue and violet colouring of the trace specifies the wavelength of the switching light of 465 nm and 365 nm, respectively.

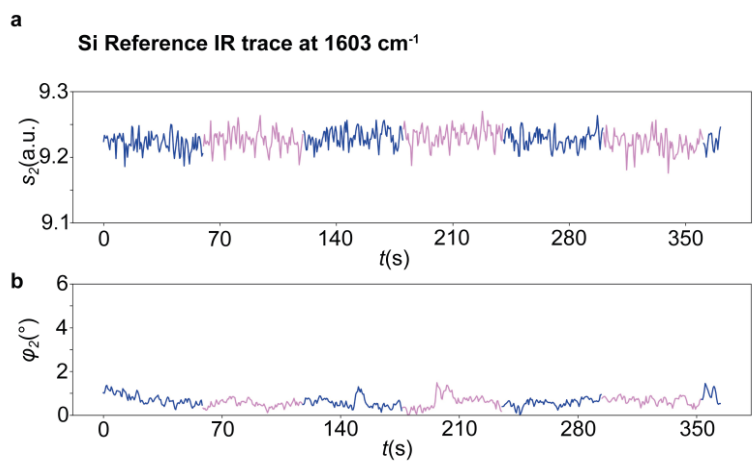

**Figure S7:** Near-field optical amplitude $s_2$ **(a)** and phase $\varphi_2$ **(b)** recorded at 1603 cm$^{-1}$ on a clean Si surface. The blue and violet colouring of the trace specifies the wavelength of the switching light of 465 nm and 365 nm, respectively.

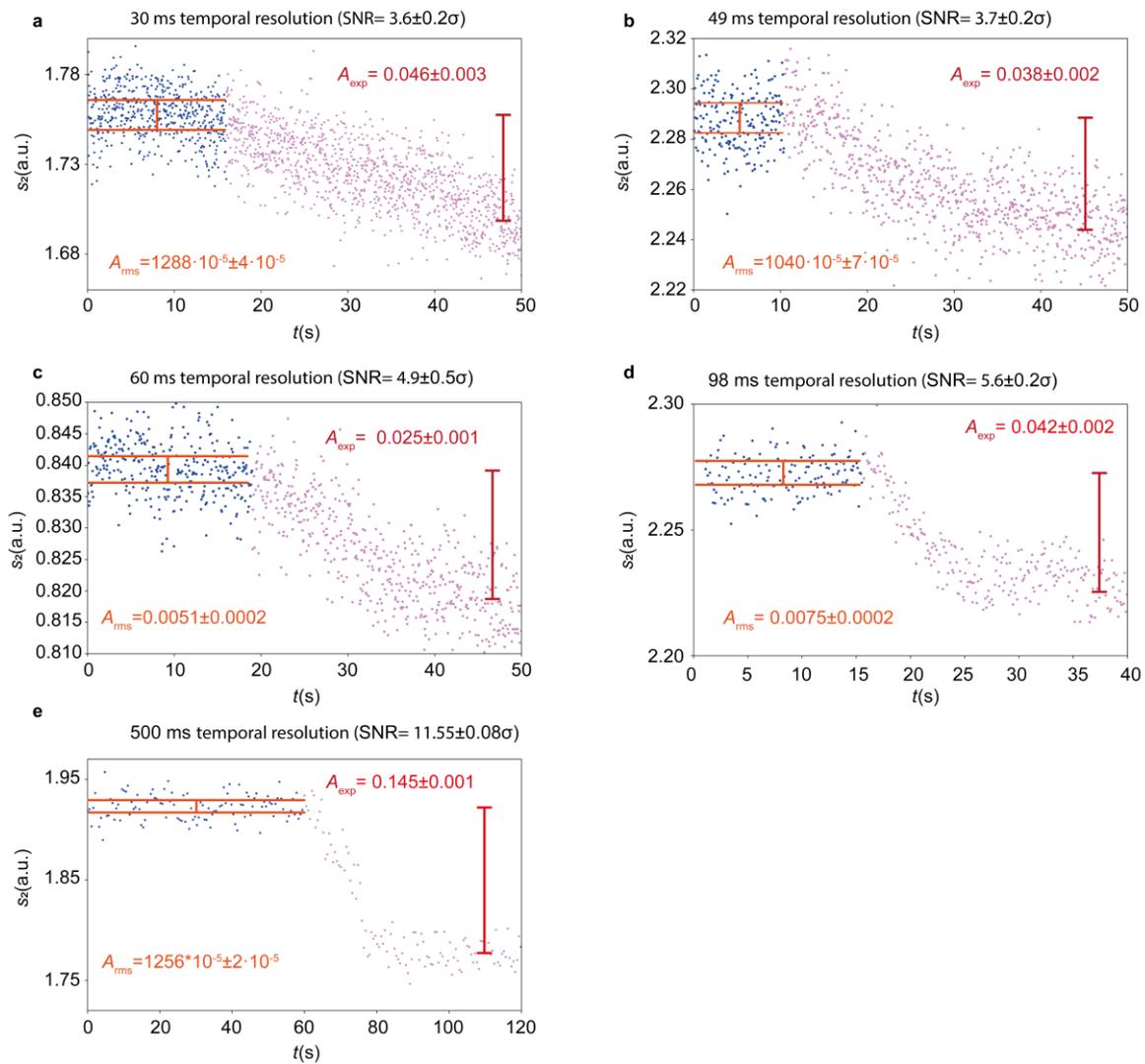

**Figure S8:** Near-field time traces of the optical amplitude $s_2$ recorded with 30 ms **(a)**, 49 ms **(b)**, 60 ms **(c)**, 98 ms **(d)** and 500 ms **(e)** temporal resolution of the *trans*-to-*cis* switching process. The recorded experimental signal values ($A_{exp}$) are marked by the red bars and the root mean-square of the noise of the steady-state signal ($A_{rms}$) before the switching perturbation is marked by the orange bars. Based on these values the signal-to-noise values for the photoswitching process have been determined and exceed for all above time-traces the commonly accepted threshold value of 3σ showing that the dynamic photoswitching process can be resolved with 30 ms temporal resolution.

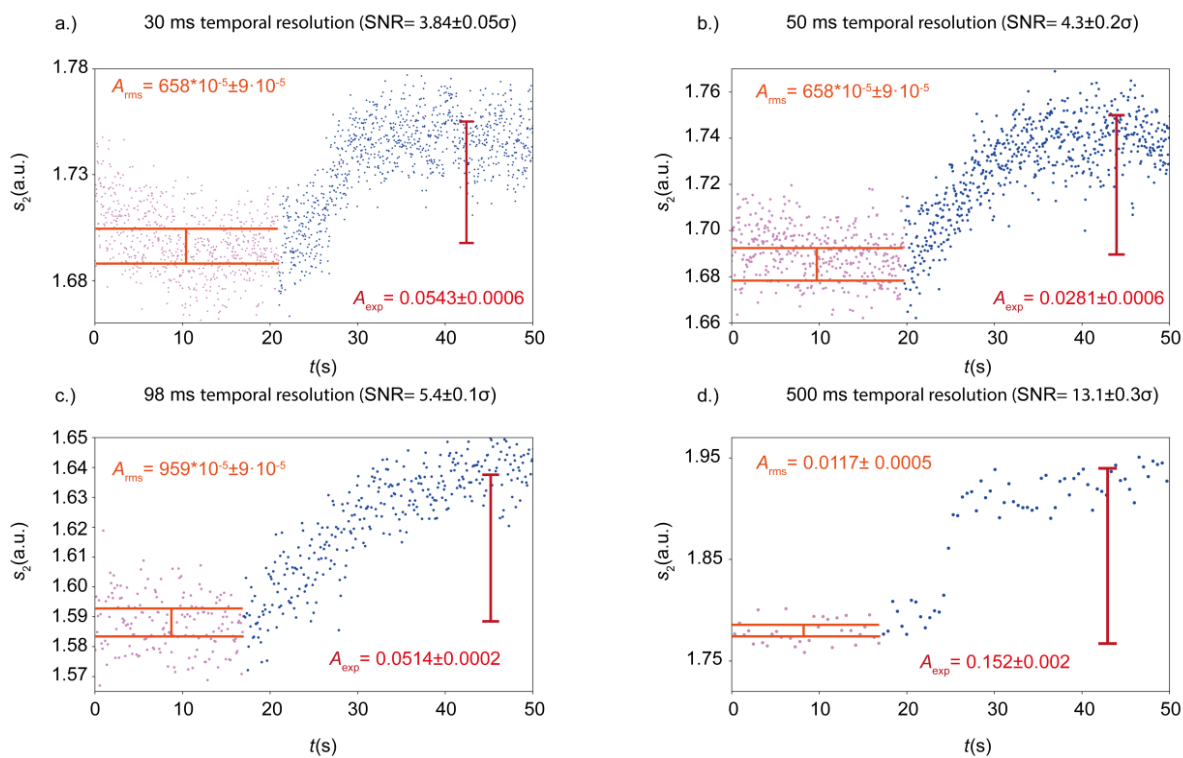

**Figure S9:** Near-field time traces of the optical amplitude $s_2$ recorded with 30 ms **(a)**, 49 ms **(b)**, 98 ms **(c)** and 500 ms **(d)** temporal resolution of the *cis*-to-*trans* switching process. The recorded experimental signal values ($A_{exp}$) are marked by the red bars and the root mean-square of the noise of the steady-state signal ($A_{rms}$) before the switching perturbation is marked by the orange bars. Based on these values the signal-to-noise values for the photoswitching process have been determined and exceed for all above time-traces the commonly accepted threshold value of 3σ showing that the dynamic photoswitching process can be resolved with 30 ms temporal resolution.

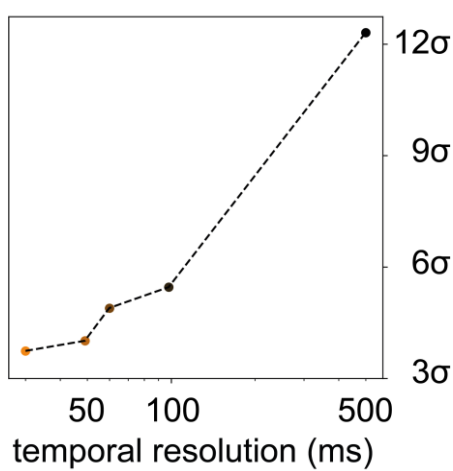

**Figure S10:** The associated signal-to-noise metrics in relationship to the recorded temporal resolution of the photoswitching dynamics shown in **Figures S8** and **S9**.

# References

1. Kaltenecker, K. J., Gölz, T., Bau, E. & Keilmann, F. Infrared-spectroscopic, dynamic near-field microscopy of living cells and nanoparticles in water. Scientific reports 11, 21860; 10.1038/s41598-021-01425-w (2021).

2. Baù, E., Gölz, T., Benoit, M., Tittl, A. & Keilmann, F. Nanoscale mechanical manipulation of ultrathin SiN membranes enabling infrared near-field microscopy of liquid-immersed samples, 2024. Preprint at https://doi.org/10.48550/arXiv.2404.01770 (2024).